\documentclass[aps,prl,twocolumn,groupedaddress]{revtex4-2}
\usepackage{amsmath}
\usepackage{amssymb}
\usepackage{graphicx}
\usepackage[colorlinks=true, allcolors=blue]{hyperref}
\usepackage[dvipsnames]{xcolor}

\newcommand{\gac}{\tilde{g}}
\newcommand{\vsound}{v}

\begin{document}
	
	\title{Wrong Signs are Alright}
 
	\author{David E. Kaplan}
	\email[]{david.kaplan@jhu.edu}
	\affiliation{Department of Physics and Astronomy, Johns Hopkins University, Baltimore, MD 21218, USA}
	
	\author{Surjeet Rajendran}
	\email[]{srajend4@jhu.edu}
	\affiliation{Department of Physics and Astronomy, Johns Hopkins University, Baltimore, MD 21218, USA}
	
	\author{Francesco Serra}
	\email[]{fserra2@jh.edu}
	\affiliation{Department of Physics and Astronomy, Johns Hopkins University, Baltimore, MD 21218, USA}
	\date{\today}
	
	\begin{abstract}
It has been shown that some Lorentz-invariant quantum field theories, such as those with higher-dimensional operators with negative coefficients, lead to superluminality on some classical backgrounds. While superluminality by itself is not logically inconsistent, these theories also predict the formation of closed time-like curves at the classical level, starting from initial conditions without such curves. This leads to the formation of a Cauchy Horizon which prevents a complete description of the time evolution of such systems. Inspired by the chronology protection arguments of General Relativity, we show that quantum mechanical effects from low energy quanta strongly backreact on such configurations, exciting unknown short-distance degrees of freedom and invalidating the classical predictions. Thus, there is no obvious low-energy obstruction to the existence of these operators. 
	\end{abstract}

	\maketitle

	\section{Introduction}
 \vspace{-10pt}
How does causality constrain effective field theories? Naively, Lorentz symmetry and locality should enforce causality. However, it has been pointed out \cite{Adams:2006sv} that even Lorentz invariant, local interactions might allow for configurations that dramatically change the standard notions of causality. Indeed, some interactions lead to faster-than-light propagation, making it possible to create classical low energy ``time-machine'' configurations, {\it i.e.} backgrounds containing closed future-directed trajectories. 

This argument suggests that faster-than-light propagation may indicate a breakdown of the low-energy description of a system, see {\it e.g.} \cite{Camanho:2014apa}, so that causality is restored in the full theory by never actually having faster-than-light excitations. Similarly, this argument supports the idea that the S-matrix should have canonical analytic properties \cite{Eden1971}. These properties depend crucially on Lorentz invariance and microcausality, which is the condition imposing that field commutators should vanish when the fields are evaluated on space-like separated regions \cite{Bros1964,Martin:1969,Streater:1989}. Both the absence of faster-than-light propagation and the analytic properties of the S-matrix lead to parametrically equivalent constraints on low-energy interactions. These constraints require the positivity of certain Wilson coefficients \cite{Pham:1985,Bellazzini:2020cot,Arkani-Hamed2021a} and the absence of Galileon-like symmetries \cite{Caron-Huot2021,Bellazzini2022,Tolley2021,Serra2023}. Therefore, according to the classical argument above, effective-field-theories (EFT) that violate positivity or display Galileon-like symmetries are in conflict with causality.

    In contrast with this classical argument, there are reasons to believe that these classical predictions of dramatic changes in the causal structure of the theory are not actually realized.  Notably, in the context of General Relativity (GR) it is known that geometries with closed null trajectories suffer from strong quantum backreaction which makes these configurations UV-sensitive \cite{Hawking:1991nk,Kim:1991mc}. It has been suggested that such backreactions may exist even in non-gravitational systems \cite{Babichev:2007dw}, with specific examples of (non-compact) configurations studied in  \cite{Tolley:2011}.
    
    Along these lines, we examine  whether quantum effects on generic classical (compact) time-machine configurations lead to a breakdown of effective-field-theory. 
	We find that similarly to the case of GR, quantum effects make these configurations strongly coupled and UV sensitive, implying that time-machines can never be described or prepared in the regime of validity of the low-energy theory. In contrast with the classical argument presented above, this means that effective-field-theories featuring negative Wilson coefficients or Galileon-like symmetries do not lead to predictions of changes to the causal structure of the theory. 

 In this work, we show that loop effects due to low-energy modes that travel on closed future-directed trajectories cause divergent backreaction on the classical configurations. The reason for these divergences is that the loop modes are on-shell{, meaning that their presence is independent on the regularization technique employed}. Even before the closed future-directed trajectory appears, slightly off-shell closed trajectories contribute to loops, causing increasingly strong backreaction. These are new, {kinematic and }configuration-dependent divergences and they cannot be removed by standard counterterms {while keeping the theory predictive}.  
	
We describe three ways in which loop effects break the low-energy description of such configurations. First, we show that the ground-state energy of these systems is affected by the new one-loop divergence in the vicinity of the closed future-directed trajectory. This divergence is triggered by low energy modes, and  in turn probes unknown UV physics. This effect is the non-gravitational analog of what is found in GR \cite{Hawking:1991nk,Kim:1991mc}. Second, we show that the dynamics of low-energy excitations around these backgrounds is strongly coupled. New loop divergences affect all operators in the effective-field-theory in the neighborhood of a closed future-directed trajectory. Third, we show that background modes of these configurations are converted into low-energy radiation at loop level. This causes a rapid decay of the configurations.
	
	These effects make it clear that these time-machine configurations are actually strongly coupled and UV sensitive. Therefore, faster-than-light propagation does not represent a pathology for the IR dynamics.
	
	In the following we describe these three effects in the context of a shift-symmetric scalar theory that allows for classical backgrounds with closed future-directed trajectories. As we discuss, our arguments are valid beyond the example we present. In light of our results, we close by discussing what UV theories could lead to faster-than-light propagation at low-energy.

\vspace{-1em}
\section{Scalar-field example}
\vspace{-0.5em}
	In this section we introduce an explicit low-energy model displaying classical time machine configurations and describe our approach to studying perturbations around these backgrounds, discussing how this generalizes to cases beyond the example at hand.

	We start by introducing the explicit model we will refer to in this work: a shift-symmetric scalar theory with superluminal propagation, as proposed in \cite{Adams:2006sv}. Consider the following Lorentz-invariant action (in signature $-+++$):
	\begin{align}\label{eq:action}
		S=\int \!\!d^4x\left\{-\frac{1}{2}(\partial\phi)^2+c_2\frac{(\partial\phi)^4}{4}+\dots\right\}\;,
	\end{align}
	where the dots indicate higher dimensional operators, negligible at low-enough energy.
	In this theory, we study scalar excitations $\varphi$ on a long wavelength background $\phi_*$: $\phi=\phi_*+\varphi$. The quadratic action for $\varphi$ will be:
	\begin{align}\label{eq:varphi}
		\!\!\!\!S_{\varphi}\!=\!-\frac{1}{2}\int \!\!d^4x\,Z^{\mu\nu}[\phi_*]\partial_\mu\varphi\partial_\nu\varphi\;.
	\end{align}
 with $Z^{\mu\nu}=(1\!-\!c_2(\partial\phi_*)^2)\eta^{\mu\nu}\!-\!2c_2\partial^\mu\phi_* \partial^\nu\phi_*\,$. 
 This action can be expressed in terms of the acoustic metric $\tilde{g}_{\mu\nu}$ defined by: $Z^{\mu\nu}=\sqrt{-\gac}\,\gac^{\mu\nu}$. This allows to rewrite the linearised equation of motion as:
 \begin{align}\label{eq:free}
\tilde{g}^{\mu\nu}\widetilde{\nabla}_\mu\partial_\nu\varphi=0\;.
 \end{align}
	where $\widetilde{\nabla}$ is the covariant derivative with respect to $\tilde{g}_{\mu\nu}$.
	This means that, under the assumption of slowly-varying background, on-shell perturbations with high enough momentum will travel following the null geodesics of the acoustic metric. Indeed, writing $\varphi=\varphi_0 e^{iS(x)}$, with $\partial_\mu S=k_\mu$, we have that in the above approximation $\tilde{g}^{\mu\nu}k_{\mu}k_\nu=0$ and $k^\mu\widetilde{\nabla}_\mu k_\nu=0$, see {\it e.g.} \cite{Weinberg:1962}. Similarly, off-shell perturbations will follow space-like or time-like geodesics of the acoustic metric.
	
	Therefore, future-directed closed trajectories will appear if $\gac_{\mu\nu}$ displays closed null geodesics.
	
	This geometric interpretation of perturbations around a background is valid beyond the example we are studying, allowing to generalize the results obtained in this work to other effective-field-theories.
	
\begin{figure}[b]
    \vspace{-2em}
		\centering
		\includegraphics[width=0.5\textwidth]{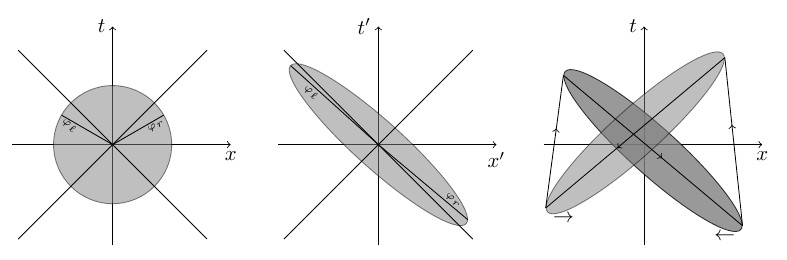}
    \vspace{-2em}
		\caption{Light cones for left moving and right moving excitations $\varphi_{\ell/r}$ in \textbf{a)} the blob's comoving frame and \textbf{b)} in a frame boosted in the positive $x$ direction with speed $\beta>1/{\vsound}$. \textbf{c)} A time-machine configuration with two blobs with high relative speed and separated by a suitable impact parameter.}
		\label{fig:boosted}
	\end{figure}

	In the theory above, we can consider a background with constant gradient $\partial_\mu\phi_*=P_\mu$, which is classically stable and satisfies the Null Energy Condition \cite{Adams:2006sv,Chandrasekaran2018}. In this setup we can see that the quartic self-interaction in Eq.~\eqref{eq:action} leads to faster-than-light propagation of $\varphi$ as soon as its coefficient is negative:
	\begin{align}
		c_2<0\;.
	\end{align}
	Indeed, writing $S=-k_\mu x^\mu=(\omega t-\Vec{k}\cdot\vec{x})$, the dispersion relation for $\varphi$ is of the form: $-\omega^2+\vec{k}^2=-2c_2(P\!\cdot \!\,k)^2$, implying a soundspeed 
	\begin{align}
		\vsound\sim1-c_2(P\!\cdot\! \hat{k})^2 \;,
	\end{align} 
	where $ \hat{k}=(1,\vec{k}/|\vec{k}|)$ and we have assumed that all scalar quantities of the form $c_2P^2$ are small.
	Given this background, we can consider a $\varphi$ excitation traveling in the $x$ direction on the background $P_\mu$. The faster-than-light dispersion relation for $\varphi$, $\left(\partial_t^2-\vsound^2\partial_x^2\right)\varphi=0\,,\,{\vsound}>1$, will transform as follows under a boost with speed $\beta$ along the $\hat{x}$ direction:
	\begin{align}\label{eq:boost}
		\!\!\!\!\!\big[(1-{\vsound}^2\beta^2)\partial_{t'}^2+2\beta(1-{\vsound}^2)\partial_{t'}\partial_{x'}\!-\!({\vsound}^2\!-\!\beta^2)\partial_{x'}^2\big]\varphi=0.\!\!
	\end{align}
	From this we can see that when $\beta>1/{\vsound}$, the boosted frame will describe some of the $\varphi$ modes propagating backwards in the time coordinate $t'$. Indeed the trajectories $x=\pm {\vsound} t$ become $x'=-\frac{{\vsound}\mp\beta}{{\vsound}\beta\mp1}t'$ in the boosted coordinates, meaning that the right-moving modes of $\varphi$ classically propagate backwards in $t'$. This is illustrated in Fig.~\ref{fig:boosted}.b.
    
	Oscillatory excitations propagating in space directions transverse to the boost in the rest frame will be described in the boosted frame by oscillatory modes with both transverse and $\hat{x}$ momentum. As noted in \cite{Adams:2006sv}, in the boosted frame purely transverse excitations have imaginary frequency. These however, will correspond to unphysical excitations with imaginary $\hat{x}$ momentum in the rest frame.

	If the entire system evolves following the time-ordering defined in the frame comoving with the blob, then physics follows a causal evolution and no issues arise. However, more complex systems will not allow for a frame in which all excitations share a global notion of time.
	For instance, one can consider two $P\neq0$ blobs traveling at high enough relative speed and spaced by an appropriate impact parameter. Such configurations allow scalar modes to travel on closed trajectories, see Fig.~\ref{fig:boosted}.c. Although no explicit solution has been constructed, in the following we will assume that closed future-directed trajectories actually appear, meaning that the acoustic metric has closed null geodesics.

\vspace{-1em}
    \section{Vacuum energy}
    \vspace{-0.5em}
	
	In this section we show that low-energy quantum excitations on a closed future-directed trajectory lead to a divergent vacuum energy, as found in the context of GR \cite{Hawking:1991nk,Kim:1991mc}. This divergence was predicted to appear beyond the context of gravity in \cite{Babichev:2007dw} and explored in specific situations in \cite{Tolley:2011}. We study this effect by computing the Green's function of the modes on a closed trajectory. Due to the periodic boundary conditions of this problem, the result is analogous to computing a Casimir energy. However, since the modes are on-shell, the trajectory is a zero-length path, leading to a new divergence.

	In the two-blobs setup described above, one can compute the expectation value of the stress-energy tensor on a point $x$ of the closed curve. Neglecting the finite background contribution $ T_{\mu\nu}[\phi]$ and using a point-splitting regularization, the expectation value can be related to a two-point function of $\varphi$:
	\begin{align}\label{eq:Tmunu}
		\langle T_{\mu\nu}(x)\rangle=\lim_{y\to x}D_{\mu\nu}\langle \varphi(x)\varphi(y)\rangle\;,
	\end{align}
	with $D_{\mu\nu}=\partial_{x^{(\mu}}\partial_{y^{\nu)}}-\frac{\eta_{\mu\nu}}{2}\partial_x\!\cdot\!\partial_{y}+O(c_2P^2)$ . 
    {Here, the point-splitting regularization serves the sole purpose of relating in a simple way the stress-energy tensor to the two-point function.}
	The Green's function for $\varphi$, $G_\varphi(x,y)$, can be computed in the WKB approximation \cite{Weinberg:1962}, or equivalently by leveraging the analogy with propagation in curved space-time \cite{DeWitt:1960,Kim:1991mc,Poisson:2011nh}. Following this second approach we define $\gamma(\tau)$ the affine $\tilde{g}-$geodesics connecting $x=\gamma(\tau_0)$ and $y=\gamma(\tau_1)$. The affine parameter $\tau$ is a scalar with dimensions of length, free up to rescaling and translation \cite{Visser:1992pz}. Each of the $\gamma$ has tangent $\ell^\mu=\frac{d\gamma^\mu}{d\tau}$, a dimensionless vector. Then $\tilde{g}^{\mu\nu}\ell_\mu \ell_\nu$ is a constant proportional to the (half) square geodesic length $\sigma_\gamma(x,y)=\frac{1}{2}(\tau_1-\tau_0)\int_{\tau_0}^{\tau_1}\tilde{g}^{\mu\nu}\ell_\mu \ell_\nu d\tau$. Without loss of generality, we can fix $\tau$ so that $\tau_1-\tau_0$ is of the order of the diameter $L$ covered by the curve in a given frame \cite{Kim:1991mc}. This choice implies that, in the same frame, the non-zero components of $\ell$ are of order $1$. With this, one can verify that the Green's function for $\varphi$ is given by the following Hadamard ansatz \cite{Hadamard:1923}:
	\begin{align}\label{eq:propag}
		G_\varphi(x,y)=\!\lim_{\epsilon\to 0^+}\!\sum_{\gamma}\!\frac{U_\gamma}{\sigma_\gamma+i\epsilon}\!+\!V_\gamma\log|\sigma_\gamma+i\epsilon|\!+\!W_\gamma\;,
	\end{align}
	with $V_\gamma\,,\,W_\gamma$ smooth functions of $x,y$, and $4\pi^2U_\gamma=\Delta_\gamma(x,y)^{1/2}$, where $\Delta_\gamma(x,y)=\det(\widetilde{\nabla}_{x}\!\widetilde{\nabla}_y\sigma_\gamma)/\!\sqrt{\tilde{g}{(x)}\tilde{g}{(y)}}$ is the van Vleck determinant of the geodesic $\gamma$ \cite{Kim:1991mc,Poisson:2011nh}. {As discussed in \cite{Kim:1991mc,Visser:1992pz}, $\Delta_\gamma=C\exp\left(-\int_{\tau_0}^{\tau_1}\widetilde{\nabla}_\mu\ell^\mu d\tau\right)$, with $C\neq0$, will be non-vanishing unless the metric infinitely defocuses the geodesics around $\gamma$, which will never be the case for the finite-size, weakly coupled backgrounds we consider.}
    In Eq.~\eqref{eq:propag}, the sum over geodesics can be understood by representing $G_\varphi$ as a sum over the eigenstates $\varphi_p$ of Eq.~\eqref{eq:free}: $G_\varphi=\sum \frac{\varphi_p(x)\varphi_p^*(y)}{\tilde{g}^{\mu\nu}p_\mu p_\nu}$, which leads to contributions from all the geodesics connecting $x$ and $y$, \mbox{see also \cite{Kim:1991mc}.} For more rigorous discussions regarding the form of Eq.~\eqref{eq:propag}, see {\it e.g.} \cite{Harte:2012uw,Casals:2012px,Kay:1996hj}.
 
    Eq.~\eqref{eq:propag} makes clear that null geodesics ({\it i.e.} on-shell modes) will give divergent contributions to the Green's function, since $\sigma_\gamma=0=\tilde{g}^{\mu\nu}k_\mu k_\nu$.
	In the limit $y\to x$, one renormalizes the two-point function by subtracting the contribution from the UV length-zero path that passes through $x$ and no other point: $G_{\varphi}^{ren}=G_{\varphi}(x,y)-\frac{1/4\pi^2}{(x-y)^2}$. This amounts to performing a wave-function renormalization of the field $\varphi$. However, the closed null geodesics will lead to a second divergent contribution that depends on the background configuration and cannot be removed by local counterterms.
	Taking into account the effect of $D_{\mu\nu}$ in the $y\to x$ limit, we see that this divergence will affect the renormalized stress-energy tensor $T^{ren}_{\mu\nu}$:
	\begin{align}\label{eq:tren}
		\langle T^{ren}_{\mu\nu}(x)\rangle \sim\lim_{y\to x}\frac{\Delta_\gamma^{1/2}t_{\mu\nu}}{\sigma_\gamma(x,y)^3}\;\gtrsim\; L^2\Lambda_{{UV}}^6\;,
	\end{align}
    since in general $\sigma_\gamma(x,y)=(\tau_1-\tau_0)\ell_\mu(x-y)^\mu$ for $y\to x$ \cite{Poisson:2011nh} and the divergence will be at best cutoff by the UV scale $\Lambda_{{UV}}$. In this equation $t_{\mu\nu}= 2(\tau_1-\tau_0)^2(\ell_\mu \ell_\nu-2\eta_{\mu\nu}\eta^{\alpha\beta}\ell_\alpha\ell_\beta+O(\sigma))$, similarly to \cite{Thorne:1992iv}. (The presence of equal powers of $\tau$ and $\ell$ makes these quantities independent on our choice of $\tau$.)
    
    Subtracting this divergence is impossible if one wishes to make the theory {predictive and} finite around the vacuum or any causal configurations. {Infact, our result in Eq.~\eqref{eq:tren} corresponds to a kinematic-dependent divergence in $n$-point correlators. This is because the localized classical backgrounds we are considering can be described as coherent states excited from the vacuum. This means that the expectation value on the classical configuration can be rewritten as a sum of vacuum expectation values of multiple field operators. The momentum space kinematics of these field operators is determined by the background, e.g. by the size and relative speed of the blobs. Therefore, the divergence in Eq.~\eqref{eq:tren} originates as a kinematic dependent divergence in the vevs that resum to the expectation value on the classical configuration. In particular, for kinematics that describe slow/distant enough blobs with no classical time-machine trajectory, all the vevs will be finite once the counterterms of the theory are fixed. Instead, changing the kinematics of the fields, one will encounter the closed on-shell trajectories and the divergence exposed above. This situation is in close analogy with the singularity encountered in GR as a result of gravitational collapse with initial conditions (kinematics) such that a black hole is formed. The same theory with initial conditions that differ only by the kinematics of the field excitations, will have no black hole formation and no singularity, making it unphysical to remove the singularity in the black hole interior by adding ad-hoc, configuration dependent counterterms.
    Another similar case is what happens in gravity-mediated scattering in 4 dimensions. There, the forward limit scattering of e.g. two particles is plagued by divergences that do not appear at finite momentum-exchange, and that cannot be consistently removed by counterterms. 
    In the case of time-machine configurations, the divergence encountered in Eq.~\eqref{eq:tren} indicates that for a system coupled to gravity or with generic self-interactions, the low-energy description breaks down and the configuration becomes UV sensitive. This invalidates the assumption that the classical time-machine configuration can be described consistently at low energy.}
	
	Even more, the divergence encountered for closed future-directed trajectories is triggered during the preparation of the configuration. Indeed, before a closed on-shell trajectory is formed, there will be closed slightly off-shell trajectories satisfying:
    \begin{align}\label{eq:offshell}
		\tilde{g}^{\mu\nu}\ell_{\mu}\ell_\nu=\lambda\;,\; \lambda\to 0\;.
	\end{align}
    This is the case since the acoustic metric will generally describe a manifold with a compactified direction that is initially space-like and becomes eventually light-like.
 
    This analysis extends to any system displaying closed future-directed trajectories in the regime of validity of the WKB approximation. Beyond this approximation, as long as the exact classical dynamics predicts closed future-directed trajectories, then divergent loop contributions from on-shell modes on these trajectories will plague the configuration.

    {In this section, our use of the point-splitting regulator has only been a tool to derive a simple relation between stress-energy tensor and two-point function. The results discussed here do not depend on the regularization technique that one would use to compute explicitly Eq.~\eqref{eq:tren}, as the divergent loop contribution can be regularized but not subtracted.}
    
	Given this strong UV dependence of time-machine configurations, one cannot conclude that faster-than-light propagation alters the causal structure of the theory.
    See for instance \cite{Costa:2005ej,McGreevy:2005ci,delaFuente:2013nba} for examples of UV descriptions of similar systems.

 \vspace{-1em}
	\section{Strong Coupling}

    In this section we show that the loop contributions studied above also produce unavoidable strong coupling for the dynamics of $\varphi$. This happens regardless of which interactions are included in the EFT and whether the system is coupled to gravity.

    The divergences that we expose in this section are due to the self-interaction of the excitations around the background. In the case of the two-blobs of scalar field \eqref{eq:action}, we have tree-level cubic and quartic self-interactions for the excitations $\varphi$: \mbox{${c}_2 \partial_\mu\phi_*\partial^\mu\varphi(\partial\varphi)^2 + \tilde{c}_2(\partial\varphi)^4/4\,$}.
    
    We start studying quantum corrections to the two-point correlator for $\varphi$ evaluated at points that are away from any closed trajectories.
    {In particular, we restrict our analysis to space-time points that can be time-ordered with respect to each other and with respect to the compact time-machine configuration.
    For instance, one could choose points that are in time-slices that do not intersect the time-machine configuration. This restriction allows to compute correlators e.g. in the in-out formalism. Indeed, despite the presence of the time-machine, one can define time-ordering through the chosen foliation of space-time and compute in-out correlators using the $\varphi$ Hamiltonian, which can be obtained considering the boosted blobs.
    Despite the separation between the chosen points and the closed trajectory, the latter still brings divergent virtual contributions to correlators, leading to strong coupling.} Focusing for simplicity on the quartic self-interaction, at one loop we have:
    \begin{align}\label{eq:loop}
        \langle\varphi{(x_1)}\varphi{(x_2)}\rangle_{{}_{1\text{loop}}}\!\!=i{c}_2\!\!
\int\!\!d^4x\partial_x G_{1,x}\partial^2_{x}G_{x,x}\partial_x G_{x,2}\;,
    \end{align}
    where $G_{i,x}\equiv G_\varphi(x_i,x)$ {- with same $\epsilon$ prescription as Eq.~\eqref{eq:propag} -} and the derivatives are contracted with a dimensionless, constant tensor that we omit. In this expression, the integral over space-time will always lead to divergent contributions in which the loop propagator is evaluated on closed on-shell trajectories: $G_\varphi(x,x)\sim 1/\tilde{g}^{\mu\nu}k_\mu k_\nu$. Other large contributions will come from slightly off-shell closed trajectories that are formed before the closed on-shell trajectories appear.
    These divergences depend on where the closed trajectories are located with respect to $x_1,x_2$, and involve both IR and UV modes. As in the case of the previous section, these effects cannot be subtracted by renormalization. Indeed the counterterms needed would make the theory infinite in any configuration with a standard causal structure. 
    Moreover, note that these effects would be present even if one had canonically normalized $\varphi$, since the background is weakly coupled and there is no appreciable screening.
    At the level of power counting the 4-volume of the closed curves scales as $L^2\sigma_\gamma$, meaning that these contributions are enhanced by $c_2/\sigma_\gamma^2$.
    
    Similar divergences from the quartic self-interaction will appear in any diagram having a loop with a single propagator. This is due to the appearance of integrals of the form $\int d^4x \dots G_\varphi(x,x)$, where the dots indicate other propagators and derivatives. Thus, all the self-interactions $(\partial\varphi)^{2n}$ will receive divergent quantum contributions.
    
    {In alternative to our approach, one could study the system by computing in-in correlators and restricting to time-slices and time intervals that do not contain the time-machine. This strategy should still probe the effects we have discussed so far.}

    These divergences at the level of the two-point function and $\varphi$ self-interactions, indicate again a large quantum backreaction which makes the classical configuration unpredictive.
    The same contributions will appear in other effective-field-theories beyond the scalar example discussed above, as long as the quantized excitations are interacting.

\vspace{-1em}
    \section{Time Machines Decay}

	In this section, we discuss instabilities that affect the classical time machine configurations, leading to their decay. As we show, loop effects make the rate of conversion of background modes into on-shell radiation divergent.

\begin{figure}[b]
		\centering
		\includegraphics[width=0.25\textwidth]{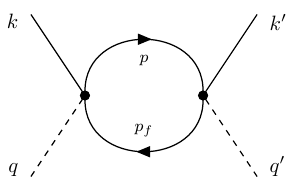}
		\caption{Conversion of background modes to radiation through a loop on the closed on-shell trajectory. The loop momenta are labeled according to their local value at the interaction vertex on the left. These values are parallel-transported along the closed geodesic.}
		\label{fig:diagram}
	\end{figure}

    The presence of a background in an interacting theory implies that excitations of the fields can extract energy and momentum from the background beyond what the acoustic metric may describe. For instance, in the case of the scalar field considered in this work, the cubic self-interaction ${c}_2 \partial_\mu\phi_*\partial^\mu\varphi(\partial\varphi)^2$ has an effective coupling ${c}_2 \partial_\mu\phi_*$ that violates coordinate translation invariance, meaning that 4-momentum is not conserved for $\varphi$ excitations.
    As a matter of fact, for a time-machine background this interaction leads to rapid production of on-shell modes radiating towards infinity.
    To see this we can consider a bubble diagram built with two cubic interactions describing the process of creation of two on-shell $\varphi$, see Fig.~\ref{fig:diagram}. 
    In order to expose a divergent contribution without making assumptions on the focusing properties of the acoustic metric, we consider a special kinematic configuration such that both the loop propagators have momentum along the closed geodesic. 
    In terms of the momenta in Fig.~\ref{fig:diagram}, this requires $p_f$ to be proportional to $p$. Requiring both the loop momenta as well as $k\,,\,k'$ to be on-shell, we see that it is possible to produce positive-energy on-shell excitations in the final state. Indeed, writing the time components of the vectors as $p_0=v|\vec{p}|\,,\,k_0=v|\vec{k}|\,,\,q_0=hv|\vec{q}|\,,\,h>0$  and defining $\vec{k}\cdot\Vec{q}=kq\cos\theta$ (with primed vectors differing by $v,h,\theta\mapsto v',h',\theta'$), we have that the process for the radiation of two on-shell modes is possible as long as:
    \begin{align}\label{eq:condition}
        \!\!\cos\theta = h\!+\!\frac{q}{2k}(1\!-\!h^2)\;,\;\cos\theta' = h'\!+\!\frac{q'}{2k'}(1\!-\!h'^2)\;.
    \end{align}
    This condition is easily satisfied {\it e.g.} if $k=q$. Momentum conservation in this process will impose $k'-q'=(q-k)_{x\to y}$, this last being the parallel transported momentum $k-q$ along the internal line labeled by $p$ from one interaction point to the other.
    {The two on-shell modes will travel to null infinity along light-like geodesics of the acoustic metric that are not trapped in the locus of closed light-like geodesics. Similarly to before, considering such modes makes it possible to define a precise time ordering and to compute in-out correlators. Even more, we can apply LSZ to the two-point correlator to derive the matrix element for the production of two on-shell modes from the classical background configuration:}
    \begin{align}
        \mathcal{A}\!=\!c_2^2\!\!\int \!\!\!d^{4}\!xd^{4}\!y \,e^{iS_{k}\!(x)}\!e^{iS_{k'}\!(y)} k_x k'_y\partial_x\phi_{\!\!{}_*}\!\partial_y\phi_{\!\!{}_*}\!(\partial_{xy}^2G_{xy})^2\!\!,\!\!
    \end{align}
    where we used the notation $k_x=\partial_x S_k(x)$.
    Similarly to the previous cases, this amplitude receives a divergent contribution from closed on-shell trajectories when Eq.~\eqref{eq:condition} is satisfied.

    If the configuration lasts an IR amount of time, as the classical evolution would predict, then the closed curves would have a four-volume $L^2\sigma_\gamma$. With this assumption, we can estimate the rate of production of $\varphi$-pairs from the background:
    \begin{align}
        \Gamma_{0\to\varphi\varphi}\sim \frac{c_2^4(\partial\phi_{\!{}_*}k^2)^4L^{16}}{\sigma_\gamma^8}\frac{1}{L^2\sigma_\gamma}\;.
    \end{align}
    Since $k\sim 1/L$, the corresponding energy loss will be \mbox{$E_{loss}\sim\frac{1}{L}(c_2/\sigma_\gamma^2)^4(\partial\phi_{{}_*}L^2)^4$}. This means that if the configuration is exactly a time-machine and on-shell modes can actually travel on closed trajectories, the lifetime of the configuration is vanishing.
    Even more, we can estimate the contribution from the emission before the closed curves become shorter than $\sigma_{min}$. The four-volume contributing to the emission will scale at least as $\sigma_{min}^2$ and the energy loss before the system reaches $\sigma_{min}$ is $E_{loss,min}\gtrsim\frac{1}{L}c_2^4(\partial\phi)^4/\sigma_{min}^4$, again divergent.

    \vspace{-1em}
	\section{Discussion}
 \vspace{-0.5em}
	In this work, we have shown how quantum effects protect the causal structure of quantum field theory at low energy, regardless of faster-than-light propagation.
	Backreaction from low-energy quanta on closed future-directed trajectories diverges, probing in turn the UV dynamics and making these configurations unpredictive. The same effect leads to very large backreaction from slightly off-shell quanta before a closed on-shell trajectory is formed, making the preparation of time-machine configurations impossible in the regime of validity of the low-energy theory. {These divergences are kinematic dependent and cannot be subtracted while keeping the theory finite and predictive, much like singularities arising in gravitational collapse or gravitational scattering.}
 
    We described three different divergent loop effects: a divergent contribution to the energy-momentum tensor, the strong coupling of excitations propagating {away from the} time machine backgrounds and the decay of the classical configuration in radiation. Although we framed our discussion in terms of a specific theory for a shift-symmetric scalar, our analysis applies as long as the closed future-directed trajectories can be described in the geometric optics approximation. Moreover, since the divergences depend uniquely on having on-shell loop contributions from closed future-directed trajectories, the same will hold beyond our present approximation. 
 
It may be surprising that such a UV back-reaction appears in regions where the naive effective field theory description should hold, namely where the backgrounds keep the irrelevant operators small.  There is a direct analogy with Cauchy horizons in General Relativity, such as at the inner horizons of spinning and charged black holes.  In those cases, the local curvature is small, yet there is a strong back-reaction both classically and quantum mechanically.  Here too, the causal structure of the theory, not the large amplitudes of the higher-dimensional operators, causes a strong back-reaction that takes the configuration outside the regime of validity of the theory.  On the other hand, backgrounds that do not alter the causal structure of the theory admit a standard effective field theory description in which naive dimensional analysis holds.

Given our results, it is interesting to reconsider the implications of faster-than-light propagation and S-matrix non-analyticity on the UV of these theories. Are there Lorentz invariant UV completions to these theories?
Microcausality violation would make it difficult to canonically quantize the fields in a Lorentz invariant theory (see {\it e.g.}, \cite{Dubovsky:2007ac}), although it might still be possible in cases without long-distance superluminal propagation \cite{Efimov:1968}. On the other hand, theories such as the DGP model \cite{Dvali:2000,Luty:2003} suggest that S-matrix non-analyticity might arise in a Lorentz-invariant setting, possibly without breaking microcausality.

Our results encourage the development of experimental methods to look for new physics related to super-luminal propagation, such as operators that have the ``wrong sign'' or approximate Galileon-like symmetries \cite{Nicolis:2008in,Pirtskhalava:2015nla,Arkani-Hamed:2002bjr,deRham:2010kj}. The discovery of such an operator would suggest the existence of UV physics that cannot simply be captured by a weakly coupled, Lorentz invariant quantum field theory. The results of \cite{Adams:2006sv} would have suggested that such experimental investigations would be fruitless due to the pathologies at low energies. We have shown that these pathologies do not arise, opening the possibility that experiment could reveal the existence of such interactions with potentially interesting implications for the full UV theory.

	\begin{acknowledgments}
		\section{Acknowledgments}
		We would like to thank 
		Nima Arkani-Hamed, Brando Bellazzini, Sergei Dubovsky, Alberto Nicolis, Alessandro Podo, Massimo Porrati, Riccardo Rattazzi, Raman Sundrum, Enrico Trincherini and Leonardo G. Trombetta 
		for useful discussions.
		This work was supported by the U.S.~Department of Energy~(DOE), Office of Science, National Quantum Information Science Research Centers, Superconducting Quantum Materials and Systems Center~(SQMS) under Contract No.~DE-AC02-07CH11359. D.E.K.~and S.R.~are supported in part by the U.S.~National Science Foundation~(NSF) under Grant No.~PHY-1818899.
		S.R. and F.S.~are also supported by the Simons Investigator Grant No.~827042 (P.I.: S.R.), and S.R. also by the~DOE under a QuantISED grant for MAGIS. 
		D.E.K.~is also supported by the Simons Investigator Grant No.~144924.
	\end{acknowledgments}
	
	\bibliography{refs}
	
\end{document}